\documentclass{rmf-d}
\usepackage{nopageno,rmfbib,multicol,times,epsf,amsmath,amssymb,cite}
\usepackage{xcolor,epsfig,amsmath,amssymb,subfigure,placeins}
\usepackage[latin1]{inputenc}
\usepackage[]{caption2}
\usepackage{graphics}
\def\({\left(}
\def\){\right)}
\def\rmfcornisa{Research OR Education of Physics \hfill\rmf\ {\bf 3} 0308123 (2022) 1-5
	\hfill MES? A\~NO?
}

\def\rmfcintilla{{\it Rev.\ Mex.\ Fis.\/} {\bf 3}  (2022)  0308123}
\clearpage \rmfcaptionstyle 
\begin{document}
	\title{  Quarkonium in a bulk viscous QGP medium 
		\vspace{-6pt}}
	\author{Lata Thakur, Yuji Hirono }
	\address{Asia Pacific Center for Theoretical Physics, \\
		Pohang, Gyeongbuk 37673, Republic of Korea }
	\author{Najmul Haque }
	\address{ School of Physical Sciences, National Institute of Science Education and Research, \\
		HBNI, Jatni 752050, India}
	\maketitle
	\recibido{day month year}{day month year
		\vspace{-12pt}}
	\begin{abstract}
\vspace{1em} 
The non-equilibrium properties 
of  quark-gluon plasma (QGP) 
have been a topic of intensive research. 
In this contribution, 
we explore the nature of heavy quarkonia 
immersed in a QGP with bulk viscosity. 
We incorporate the 
bulk viscous effect through the deformation 
of the distribution functions of thermal quarks and gluons, with which the color dielectric permittivity can be computed. 
We use the color dielectric permittivity to 
compute the heavy quark potential 
inside a bulk viscous plasma 
and solve the Schr\"odinger equation 
using the potential 
to obtain the physical properties such as binding energies and decay widths. 
We discuss the effect of the bulk viscous correction on the quarkonium properties and 
the melting temperatures.

\vspace{1em}
\end{abstract}
\keys{Quark-gluon plasma, heavy quarkonia, bulk viscosity \vspace{-4pt}}
\pacs{  14.40.Pq, 12.38.Mh, 21.65.Qr, 25.75.-q, 11.25.Tq, 47.17.+e    \vspace{-4pt}}
\begin{multicols}{2}

\section{Introduction}
Heavy quarkonia  have been one 
of the  key probes of  quark-gluon plasma (QGP) 
created in heavy-ion collisions. 
The suppression of quarkonium states 
was predicted to indicate 
the formation of  QGP~\cite{Matsui:1986dk}. 
This suppression occurs because of the Debye screening, which is reflected in the deformation of the real part of a heavy quark potential ~\cite{Matsui:1986dk}, 
and the in-medium transitions, 
which are encoded in the imaginary part of the  potential~\cite{Laine:2006ns,Brambilla:2008cx}.

In this contribution, we examine the effect of non-equilibrium bulk viscosity on the properties of heavy quarkonia~\cite{Thakur:2020ifi}.
Recent works show that the bulk viscosity of the QCD medium is enhanced near the critical point~\cite{Kharzeev:2007wb,Karsch:2007jc,Moore:2008ws,Noronha-Hostler:2008kkf,Ryu:2015vwa,Monnai:2016kud}, 
which may give rise to observable effects 
in the beam energy scan program~\cite{Bzdak:2019pkr}. 
The purpose of this study is to test 
how sensitive heavy quarkonia are 
to the bulk viscous nature of a QGP fluid. 
If they indeed are sensitive, 
there is a possibility to use heavy quarkonia 
as a probe of the changes of 
the bulk viscosity of a QGP.

The outline of the computational procedure is as follows. 
To encode the bulk viscous effect, 
we first deform the thermal distribution 
of quarks and gluons,
and compute the color dielectric permittivity 
of the medium using the Hard Thermal Loop (HTL) 
approximation at one loop~\cite{Du:2016wdx,Thakur:2020ifi}.
Using the modified dielectric permittivity, 
we compute the heavy quark potential 
of a bulk viscous medium. 
We solve the Schr\"odinger equation 
using the real part of the obtained heavy quark potential
and compute the physical properties of 
heavy quarkonium states 
such as binding energies and decay widths.

\section{Color dielectric permittivity of a bulk viscous  medium}

Here, we outline the computation of the dielectric 
permittivity of a bulk viscous medium. 

We incorporate the non-equilibrium bulk viscous
effect by modifying the distribution 
functions of thermal quarks and gluons, 
\begin{equation}
	f({\bf p})=f_{\rm eq}(p)+\delta_{\rm bulk} f(\bf p)  ,
\end{equation}
where $ f_{\rm eq}(p) $ is the equilibrium distribution and $\delta_{\rm bulk} f(\bf p) $ is 
the bulk viscous correction, 
which we parametrize in the following 
way~\cite{Du:2016wdx}, 
\begin{equation}
	\delta_{\rm bulk} f (p)=\left(\frac{p}{T}\right)^{a}\Phi\  f_{0}(p) \left(1\pm f_{0}(p)\right),
	\label{eq:f-bulk}
\end{equation}
where  $ +(-) $ indicates the Bose (Fermi) distribution. 
The parameters $\Phi$ and $a$ control the strength of the bulk viscous correction. 

Using the deformed distribution function \eqref{eq:f-bulk}, the retarded self-energy in the presence of bulk correction becomes 
\begin{equation}
	\begin{split} 
		\Pi_{R}(K)
		&=\widetilde {m}_{D,R}^2
		\left(\frac{k^{0}}{2k}\ln\frac{k^{0}+k+i\epsilon}{k^{0}-k+i\epsilon}-1\right)  , 
		\label{eq:pir-bulk}
	\end{split}
\end{equation}
where 
$K = (k^0, \mathbf k)$ is the four momenta, 
$\widetilde {m}_{D,R}^2 \equiv m^{2}_{D}+\delta m^{2}_{D,R}$,
and $m^{2}_{D}$ is the equilibrium Debye mass. 
The correction from the bulk viscosity $\delta m^{2}_{D,R}$ 
is written as 
\begin{equation}
	\delta m^{2}_{D,R}
	=
	\frac{g^2T^2}{6}\left[2N_{c}
	c^{g}_{R}(a)\Phi
	+N_{f} \left(1+\frac{3 \tilde \mu^{2}}{\pi^{2}}\right) 
	c^{q}_{R}(a,\tilde \mu)\Phi
	\right] ,
	\label{mDRtot}
\end{equation}
where 
$ c^{q}_{R}(a,\tilde \mu) $ and $ c^{g}_{R}(a )$ are  dimensionless quantities~\cite{Du:2016wdx}. 
The symmetric self-energy can also be computed in a similar manner, 
and again the effect of the bulk viscous correction 
enters through the modification of the Debye mass, 
$\widetilde m^2_{D, S} =  m^2_D + \delta m^2_{D, S}$. 
Using the self energies, 
one can compute the retarded, advanced and symmetric propagators. 
Based on these quantities, 
the dielectric permittivity $ \varepsilon(k) $ is
expressed 
using the modified Debye masses as 
\begin{equation}
	\varepsilon^{-1}(k) = 
	\frac{k^2}{k^2 + \widetilde{m}^2_{D,R}}
	- i  
	\frac{\pi T k  \, 
		\widetilde{m}^2_{D,S}  
	}{ (k^2 + \widetilde{m}^2_{D,R})^2} . 
	\label{eq:epsilonphi}
\end{equation}
\section{In-medium heavy quark potential }\label{sec:pot}

The color dielectric permittivity \eqref{eq:epsilonphi} encodes the bulk viscous effect of the medium. 
In the spirit of the linear-response theory\cite{Thakur:2013nia,Thakur:2016cki}, 
we compute the in-medium heavy quark potential by 
modifying (the Fourier transform of) the Cornell potential with the color dielectric permittivity. 
The real part of the modified potential reads 
\begin{equation}
	\begin{split} 
		{\rm Re\,} V(r)
		&= 
		\int \frac{d^3\mathbf k}{{(2\pi)}^{3/2}}
		(e^{i\mathbf{k} \cdot \mathbf{r}}-1)
		V_{\rm Cornell} (k)
		{\rm Re\,} \varepsilon^{-1}(k)  \\
		&= 
		-\alpha \, 
		\widetilde{m}_{D,R}
		\left(\frac{e^{-\widetilde{m}_{D,R}\, r}}{\widetilde{m}_{D,R}\, r}+1\right) 
		+ 
		\frac{2\sigma}{
			\widetilde{m}_{D,R}}\\
		&\times	\left(\frac{e^{-\widetilde{m}_{D,R}\,{r}}-1}{\widetilde{m}_{D,R}\,r}+1\right) ,
		\label{eq:pot-real}
	\end{split}
\end{equation}
and the imaginary part is
\begin{equation}
	\begin{split} 
		{\rm Im\,} V(r) 
		&=\int \frac{d^3\mathbf k}{{(2\pi)}^{3/2}}
		(e^{i\mathbf{k} \cdot \mathbf{r}}-1)
		V_{\rm Cornell}(k) {\rm Im\,} \varepsilon^{-1}  (k)  
		\\
		&= 
		-\alpha \lambda T 
		\, \phi_2  (\widetilde{m}_{D,R}\, r)
		- 
		\frac{ 
			2\sigma T 
			\lambda 
		} {
			\widetilde{m}^2_{D, R} 
		}
		\, 
		\chi( \widetilde{m}_{D, R}\, r ) ,
		\label{eq:pot-imaginary}
	\end{split}
\end{equation}
where 
$\lambda = \widetilde m^2_{D, S}/\widetilde m^2_{D, R}$, 
$V_{\rm Cornell} (k) $ is the Fourier transform of the Cornell potential, 
and $ \alpha \equiv C_{F} \alpha_{s} $ with 
$C_{F}=(N_c^2-1)/2N_c$. 
We take the string tension to be 
$ \sigma=(0.44 \, {\rm GeV})^{2}$, 
and $\alpha_{s}$ 
to be the one-loop coupling constant~\cite{Thakur:2020ifi}. 
\end{multicols}
\begin{figure}[tb]
\subfigure{
	\hspace{-0mm}\includegraphics[width=7.7cm]{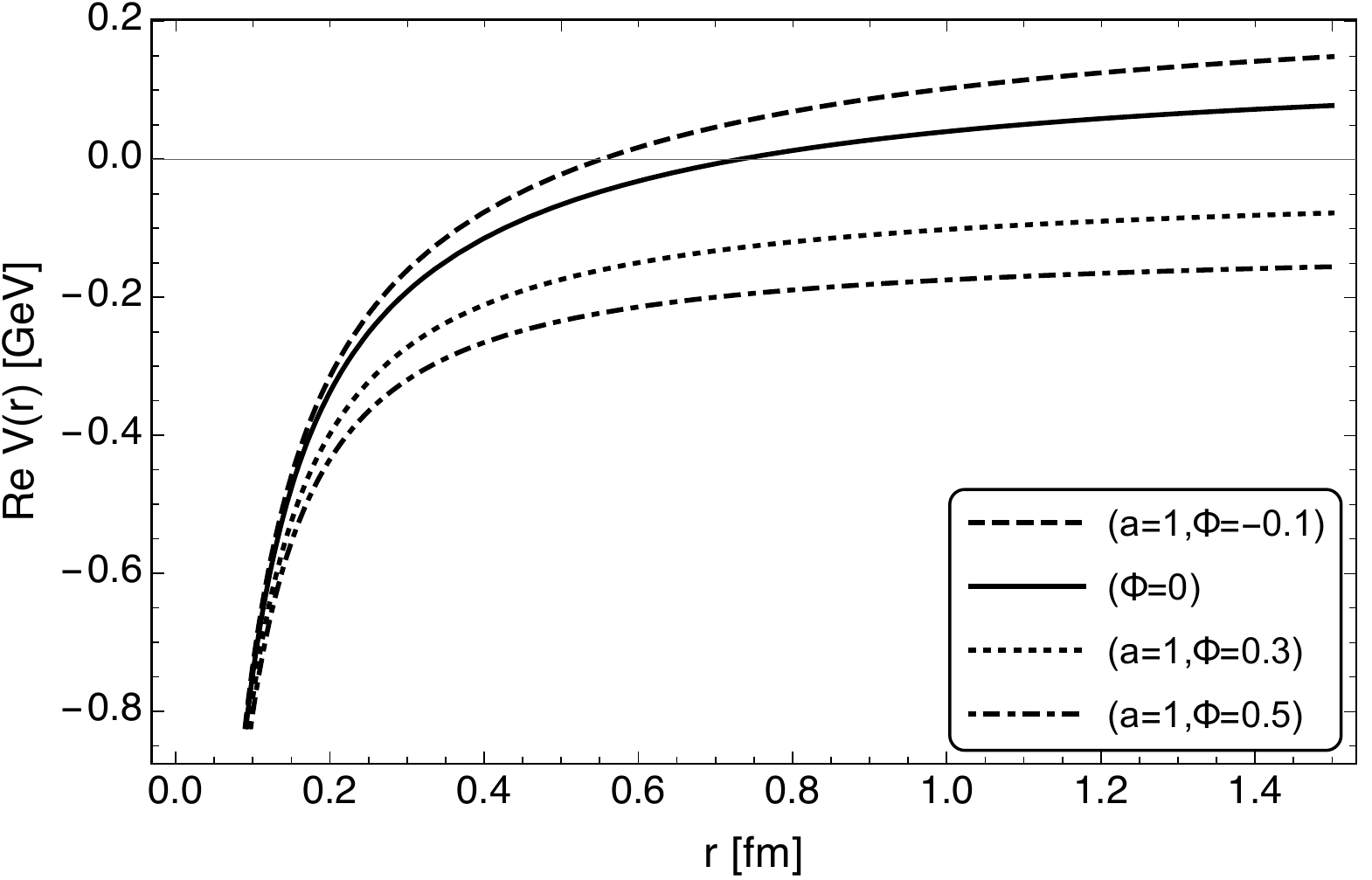}} 
\hspace{1cm}
\subfigure{
	\includegraphics[width=7.7cm]{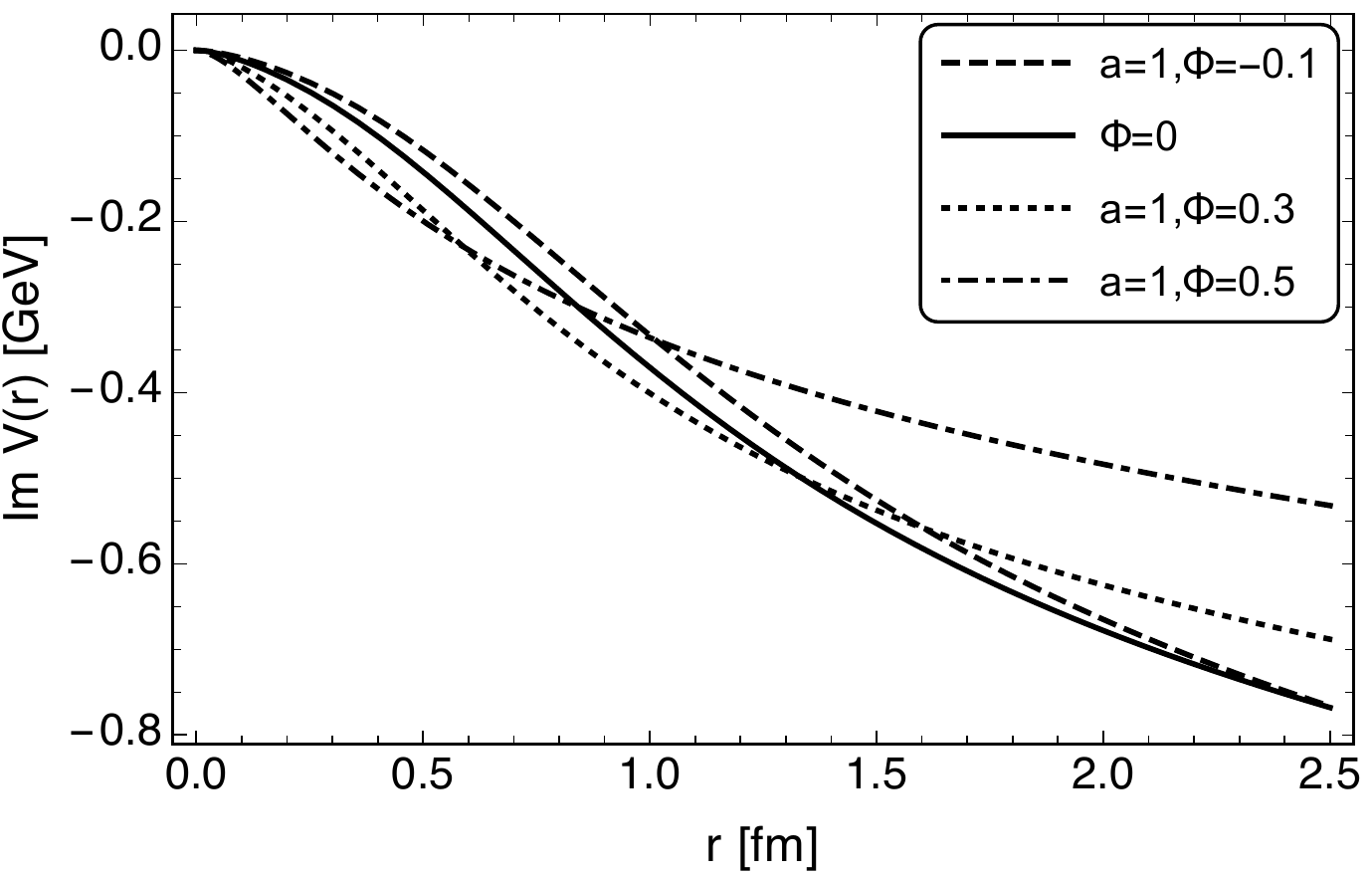}}
\caption{
	Real part (left)
	and imaginary part (right) 
	of the modified potential 
	as a function of the distance $r$ 
	at $  T=0.3  $ GeV.
}
\label{pot_re_rmD}
\end{figure}
\medline
\begin{multicols}{2}
In Fig.~\ref{pot_re_rmD}, 
we plot the real and imaginary parts of the 
modified potential. 
Different lines correspond to different values of the parameter $\Phi$. 
The real part of the potential becomes flattened 
for a larger value of $ \Phi $ 
because of a stronger screening effect (larger Debye mass).  
The influence of the bulk viscosity 
on the imaginary part is more nontrivial. 
Its magnitude, $|{\rm Im\,} V|$, shows an enhancement in the small $r$ region and 
a suppression in the large $r$ region in the presence of $ \Phi >0$.

\section{Heavy quarkonia in a bulk viscous medium} 

Let us now study the properties of 
heavy quarkonia in a bulk viscous medium. 
Using the modified potential (\ref{eq:pot-real}), 
we solve the Schr\"odinger equation 
for the radial wave function, 
\begin{small} 
	\begin{equation}
		\begin{split}
			&- \frac{1}{2 m_q}
			\left( 
			\psi'' (r) + \frac{2}r \psi' (r)
			- \frac{ \ell (\ell+1)}{r^2} \psi (r)
			\right)
			+ 
			{\rm Re\,} V(r) \,  \psi  (r) \\
			& \quad \quad = 
			\epsilon_{_{n \ell}} \,  \psi(r) ,
		\end{split}
		\label{eq:schro}
	\end{equation}
\end{small} 
where $m_q$ is the reduced mass of the quarkonium system. 
We take $m_q = 1.25 /2 \,{\rm GeV} $ for $c \bar c$ 
and $m_q = 4.66 /2 \, {\rm GeV} $ for $b \bar b$. 
%
Equation \eqref{eq:schro} is solved 
numerically to obtain 
the wave functions and eigenvalues.
The binding energy, $ E_{\rm B}$, 
and the decay width, $\Gamma$, 
are given by 
\begin{align} 
	E_{\rm B} &= {\rm Re\,} V(r \to \infty) - \epsilon_{_{n \ell}} , \\
	\Gamma
	&= - \langle \psi | {\rm Im\,} V(r) | \psi \rangle .
\end{align}

\begin{figure}[tb]
\begin{minipage}[b]{0.5\linewidth}
	\centering
	\includegraphics[width=7.8cm]{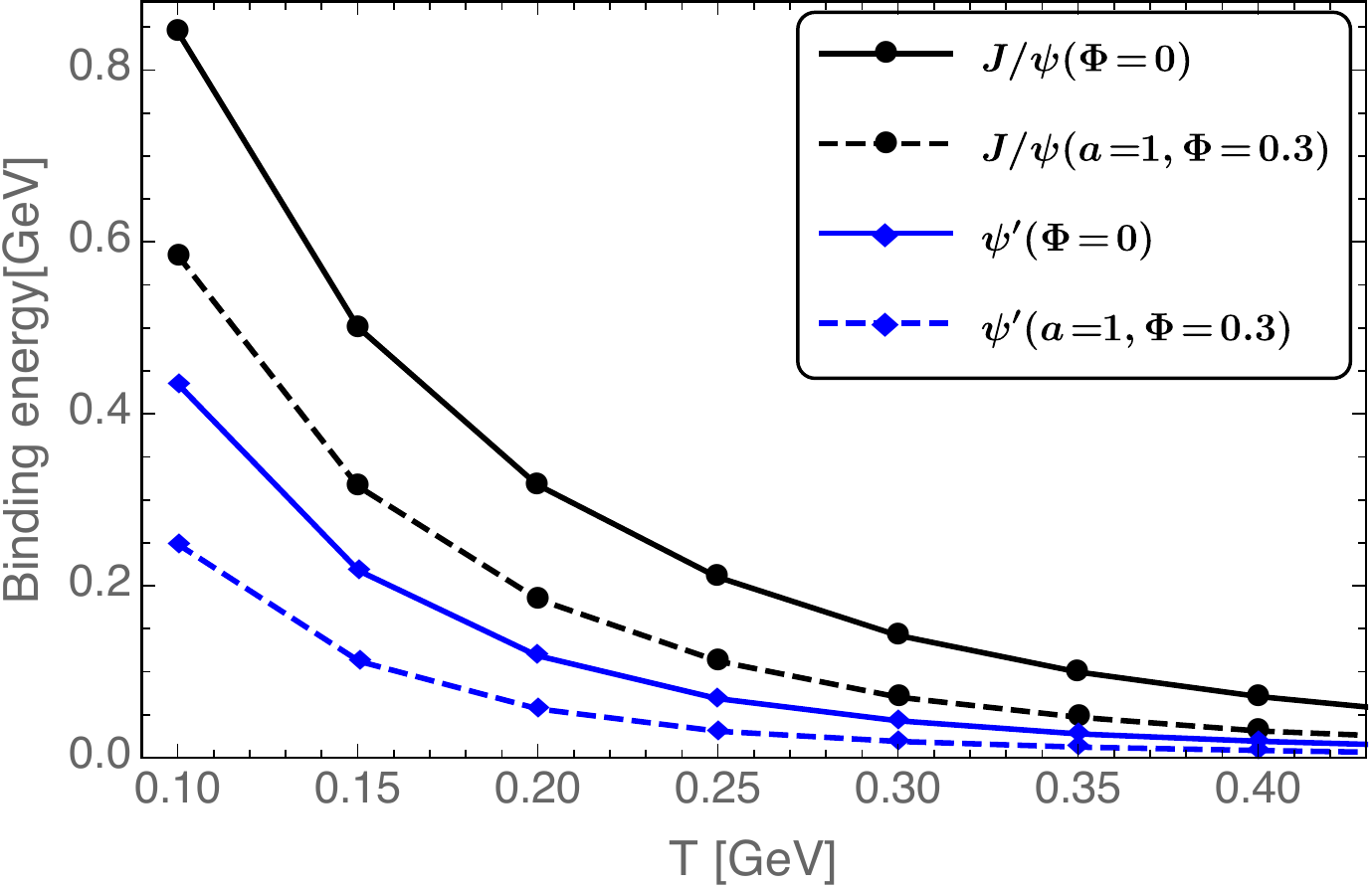}
\end{minipage}
\begin{minipage}[b]{0.5\linewidth}
	\centering
	\includegraphics[width=7.8cm]{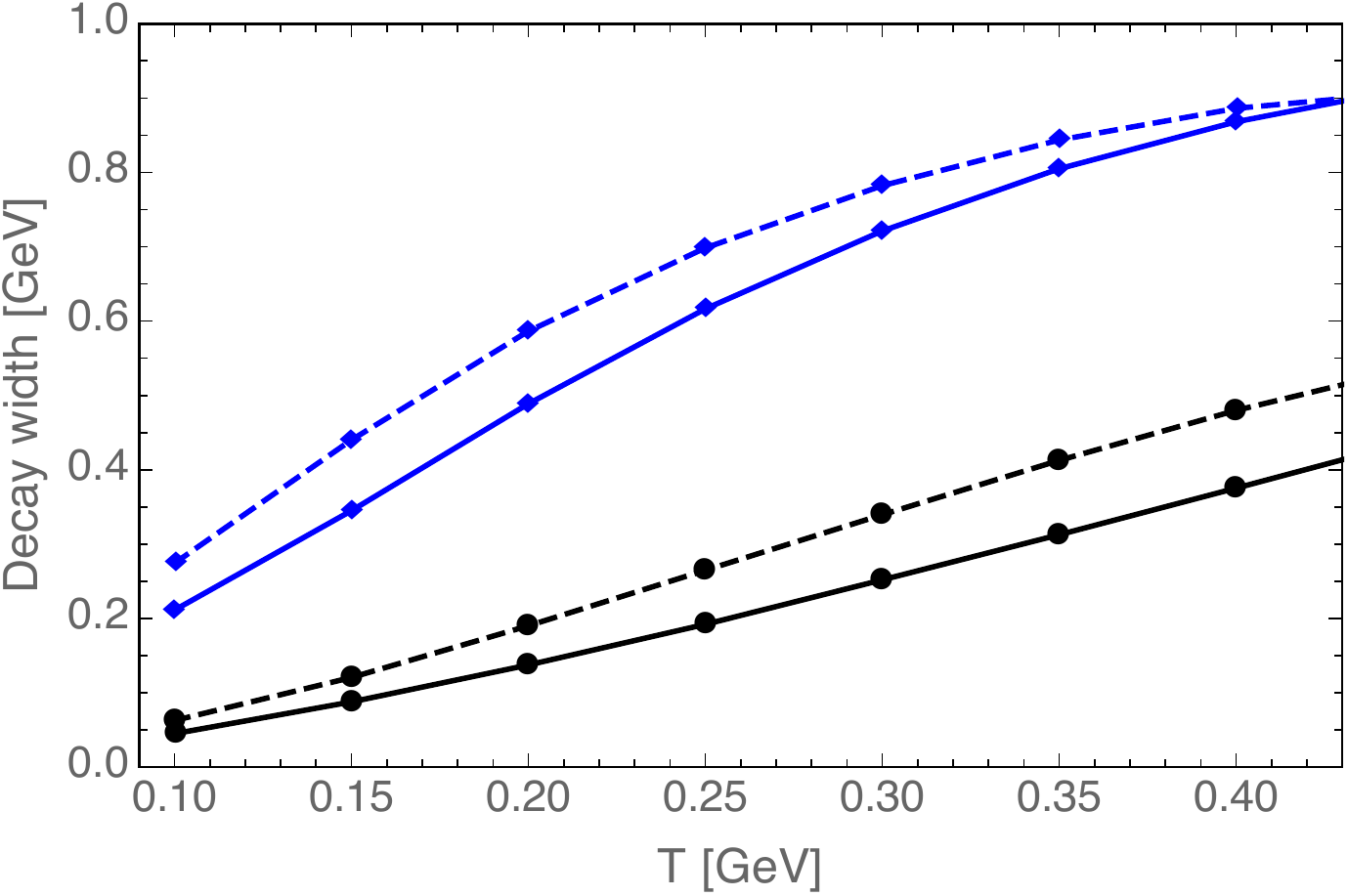}
\end{minipage} \\
\begin{minipage}[b]{0.5\linewidth}
	\centering
	\includegraphics[width=7.8cm]{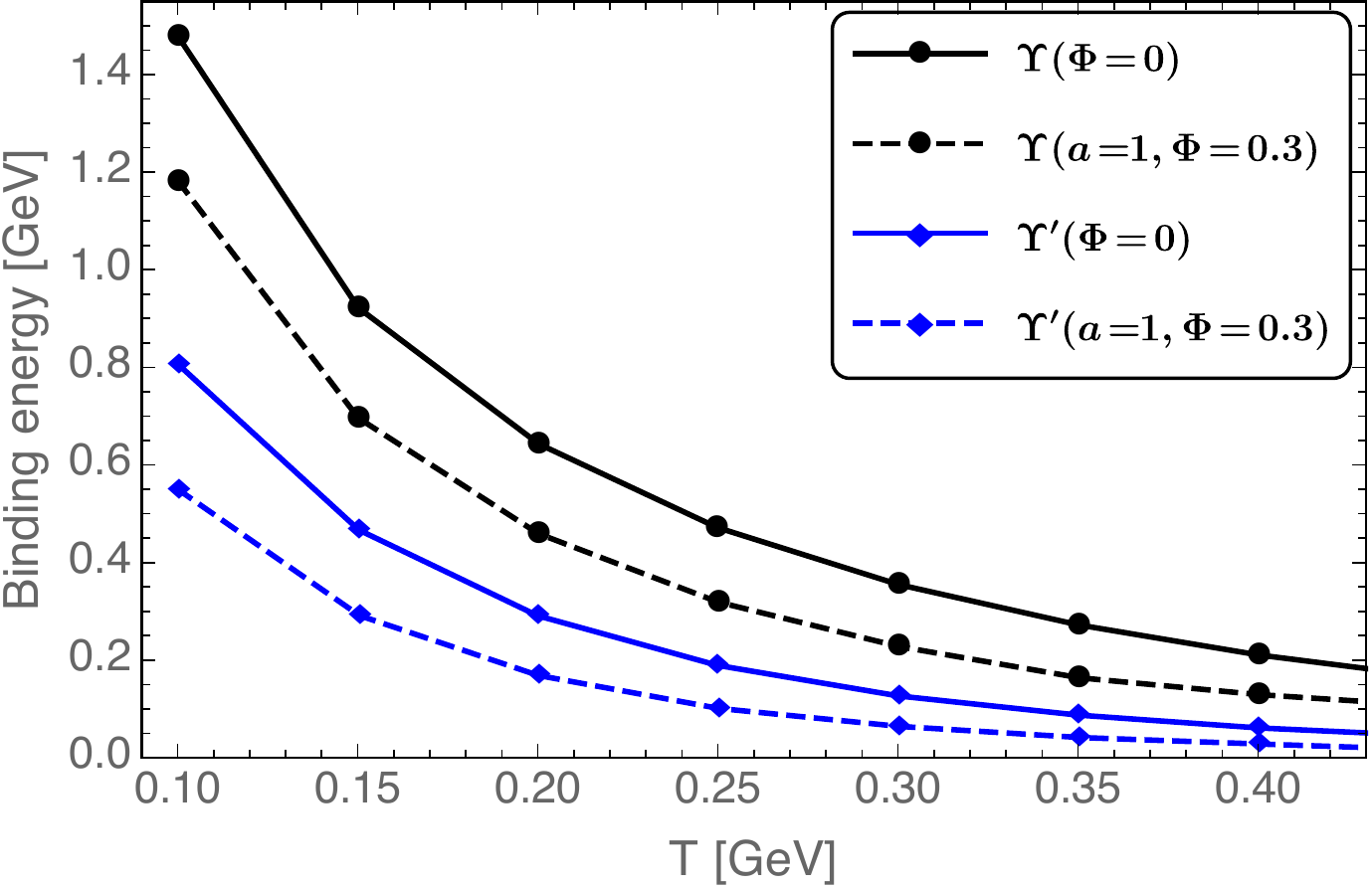}
\end{minipage}
\begin{minipage}[b]{0.5\linewidth}
	\centering
	\includegraphics[width=7.8cm]{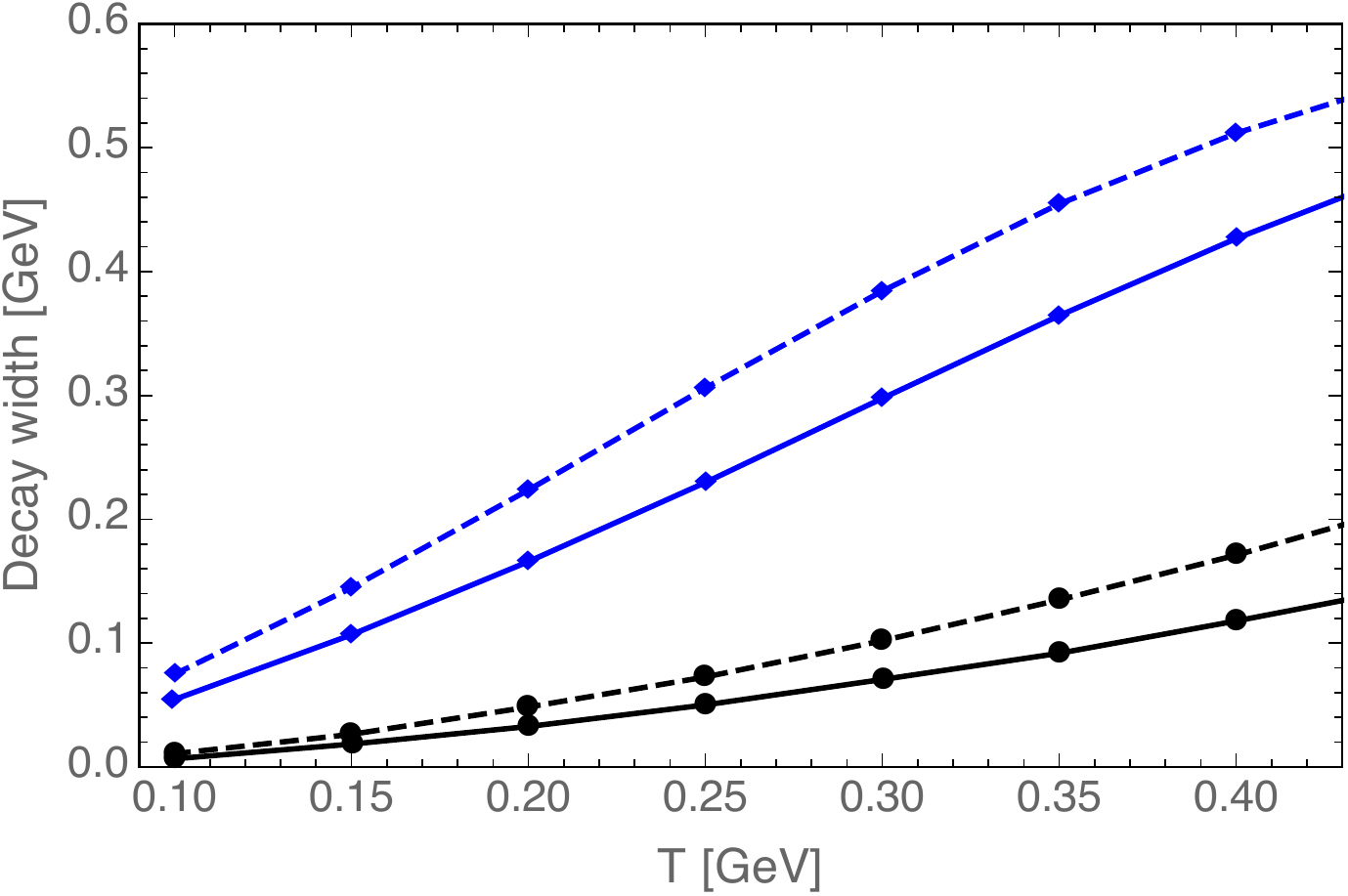}
\end{minipage} 
\caption{
	Binding energies (left)
	and decay widths (right) for 
	charmonium (top) and bottomonium (bottom) states as a function of temperature  with (dashed lines) and without (solid lines) bulk viscous corrections. 
}\label{fig:e-gamma-bulk}
\end{figure}

In Fig.~\ref{fig:e-gamma-bulk}, 
we plot the binding energies  (left) and decay widths (right)
for charmonium (top)  and bottomonium (bottom) states. 
The solid and dashed lines correspond to 
cases with and without bulk viscous corrections, 
respectively. 
We observe that the binding energy is a decreasing function of the bulk viscous correction, which is because of a stronger screening effect. 
The decay width shows 
an enhancement as a function of $ \Phi $. 
\end{multicols}
\begin{figure}[tb]
\centering
\includegraphics[width=8.5cm]{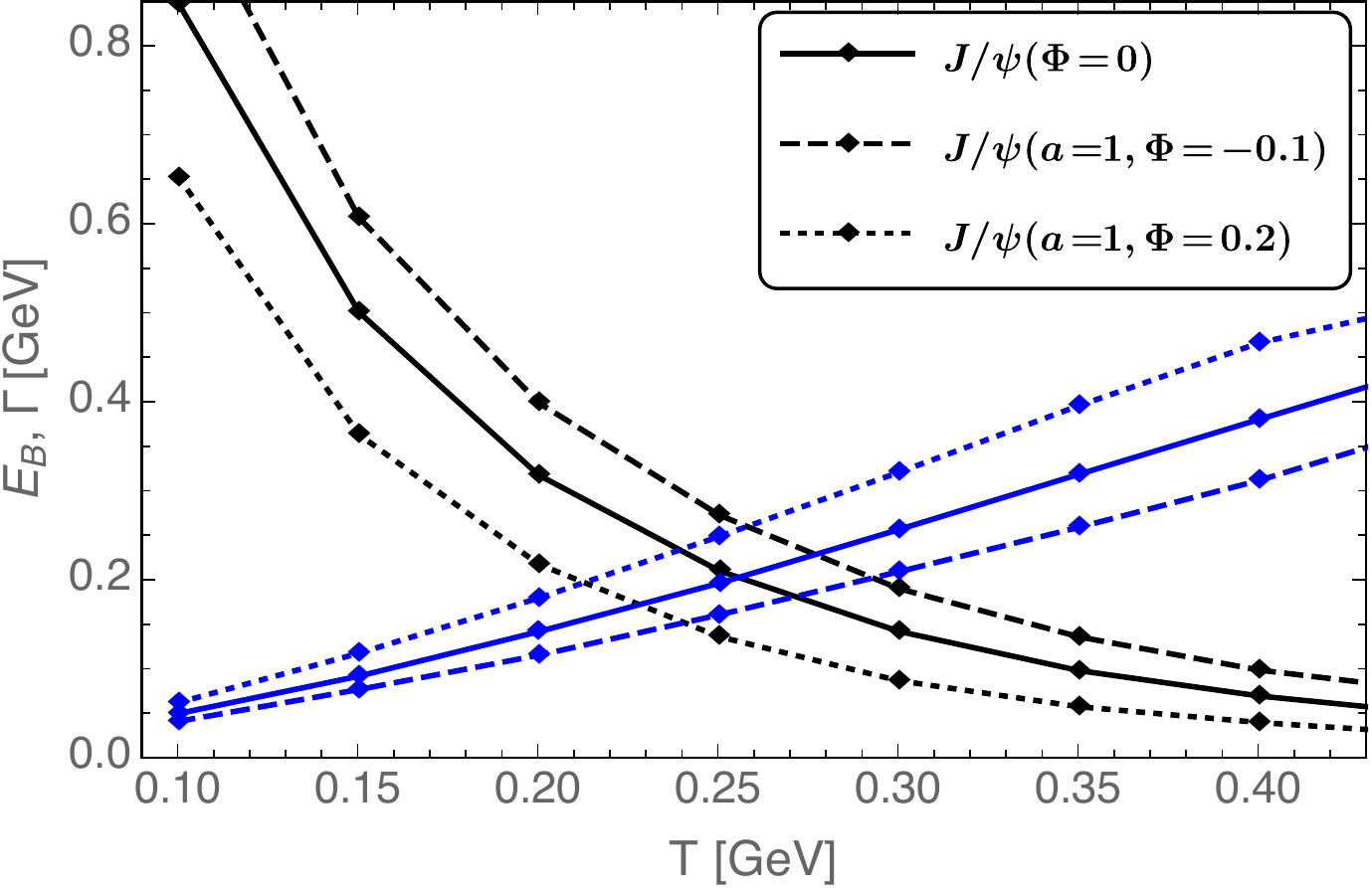}
\caption{
	Decay widths (blue lines) and binding energies (black lines)
	of $J/\psi$ for different values of $\Phi$.
	The intersection of binding energy and decay width for the same value of $\Phi$ indicate the melting temperature.
}
\label{fig:jpsi-e-gamma}
\end{figure}

\medline
\begin{multicols}{2}
In Fig.~\ref{fig:jpsi-e-gamma}, 
we plot the 
binding energies (black line) and decay widths (blue line) of $J/\psi$ states for different values of $\Phi$ at $a=1$. 
The melting temperatures $T_{\rm melt}$ of  quarkonium states
are estimated based on a standard criterion 
that the binding energy intersects with the decay width at the melting temperature, 
$E_{\rm B}(T_{\rm melt}) = \Gamma(T_{\rm melt})$.  
In Table~\ref{tab:meltingT}, 
we compare 
the melting temperatures computed in the current computation for $\Phi=0$ 
with the results from Ref.~\cite{Lafferty:2019jpr}, 
which are based on the extraction of the potential from lattice QCD data.
The current model reproduces 
the former results, at least qualitatively. 
To make an estimate of a systematic dependence, 
we varied the value of parameter $\Lambda$ 
in the one-loop coupling. 
The upper and lower number in the table correspond to 
$\Lambda = 1/2 \times 0.176 \, {\rm GeV}$ 
and 
$\Lambda = 2 \times 0.176 \, {\rm GeV}$, respectively.

\begin{figure}[tb]
\subfigure{
	\includegraphics[width=7.8cm]{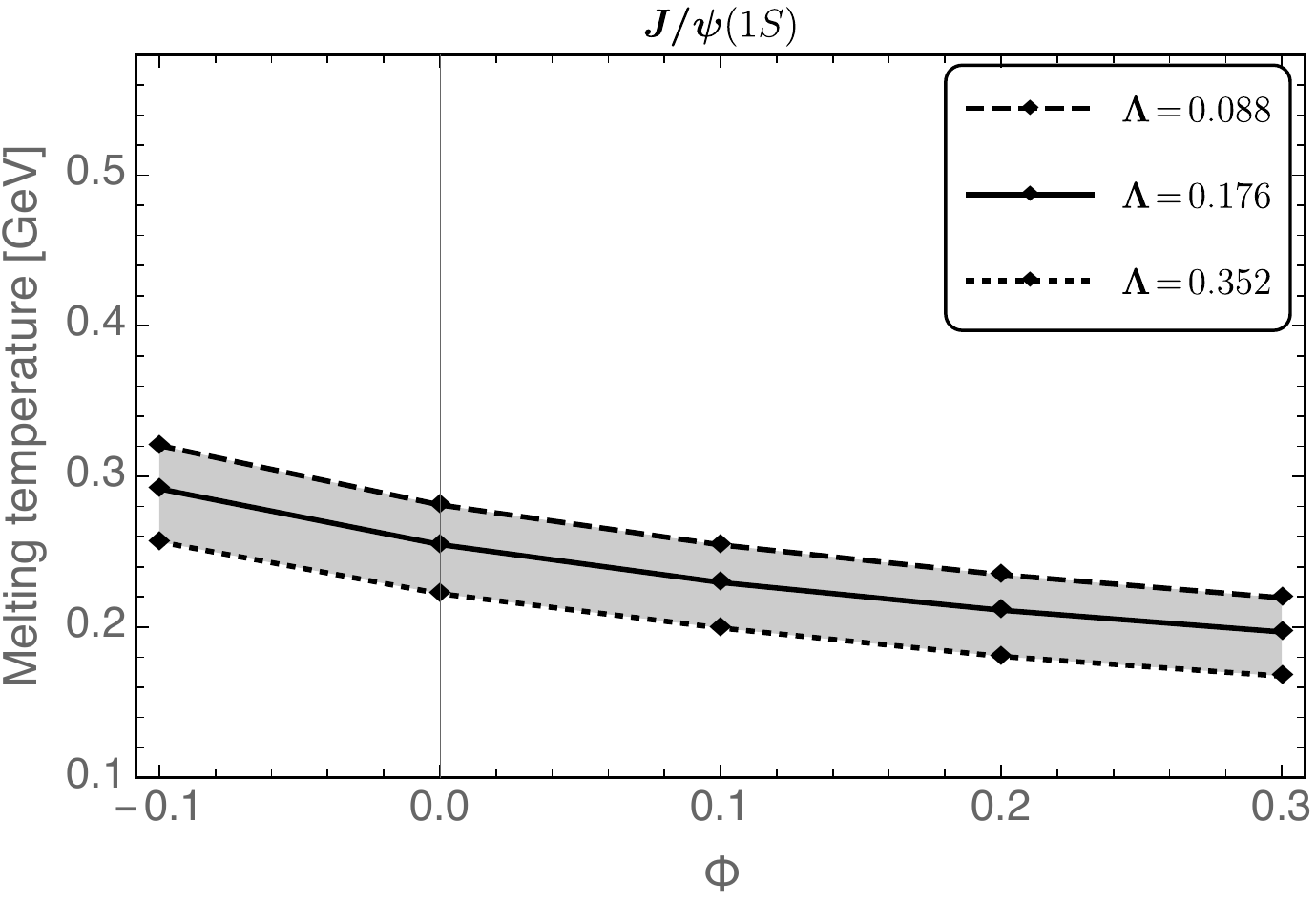}
}
\hspace{1cm}
\subfigure{
	\includegraphics[width=7.8cm]{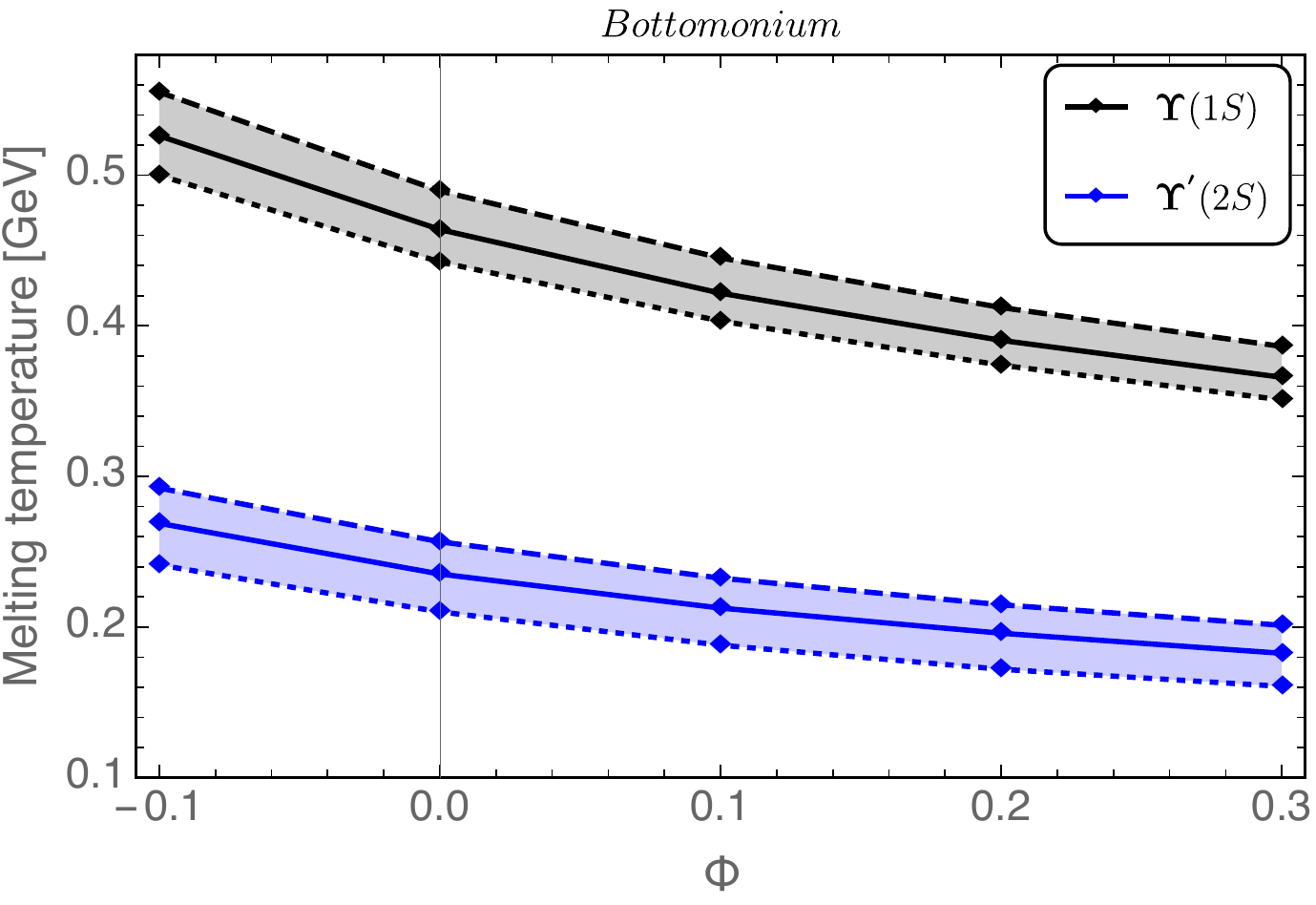}
}
\caption{
	Melting temperature
	of charmonium ($J/\psi$) (left) and bottomonium ($\Upsilon$, $\Upsilon'$) (right) 
	as a function of the bulk viscous parameter 
	$\Phi$ with $a=1$. 
}
\label{fig:t-melt-c-b}
\end{figure}

Now, let us see the influence of the bulk viscosity 
on the melting temperatures. 
Fig.~\ref{fig:t-melt-c-b} shows the melting temperatures 
of the  $J/\psi$, $\Upsilon$ and $\Upsilon'$ states
as a function of $\Phi$. 
We observe that the melting temperatures of the quarkonium states decrease 
slightly as a function of $\Phi$. 
%
%
%
If there is a sudden enhancement in the bulk viscous effect at a certain collision energy, 
the behaviour of observables related to heavy quarkonia
might show anomalous behaviour. 
It would be interesting to look at those observables 
and their collision energy dependence 
on the beam energy scan program, 
since the bulk viscous effect is expected 
to become larger near the critical point. 
\end{multicols}
\begin{table}
\centering
\begin{tabular}{ c  c   c }
	\hline
	\hline
	& $\quad T_{\rm melt}$ [GeV]  $\quad$ &
	$\quad\quad$ Ref.~\cite{Lafferty:2019jpr}$\quad\quad$\\
	\hline
	$J/\psi$              &
	$0.254^{+0.027}_{-0.032}$  &$0.267^{+0.033}_{-0.036}$\\
	$\Upsilon$        
	& $0.464^{+0.026}_{-0.022}$   &$0.440^{+0.080}_{-0.055}$\\
	$\Upsilon^\prime$  
	& $0.235^{+0.025}_{-0.022}$  & $0.250^{+0.050}_{-0.053}$\\
	\hline\hline
\end{tabular} 	
\caption{
	Melting temperatures 
	of $J/\psi$, $\Upsilon$ and $\Upsilon^\prime$ in the absence of bulk viscous correction ($\Phi=0$) 
	computed in the current model (left column). 
	The right column shows 
	the results from Ref.~\cite{Lafferty:2019jpr} based on the lattice QCD.} 
\label{tab:meltingT}
\end{table}

\medline
\begin{multicols}{2}

\section{Conclusion}
In this contribution, 
we have discussed the effects of
bulk viscous corrections on 
the properties of quarkonium states. 
We computed the color dielectric permittivity
of a bulk viscous QGP 
and computed a modified heavy quark potential. 
We have discussed how the bulk viscous correction 
deforms the real and imaginary parts of the potential.
Using the modified potential,
we solved the Schr\"{o}dinger equation to obtain the wave functions, binding energies and decay widths of the quarkonium states. 
We found that 
the binding energy decreases 
and decay width increases 
as a function of $\Phi$ for a fixed temperature. 
We also found that the melting temperatures
of the quarkonium states are reduced for $\Phi > 0$. 
%
We expect that the bulk viscous effect is 
enhanced near the critical point
and it would be interesting 
to look at 
how the physical 
observables such as $R_{AA}$ behaves 
as a function of collision energy. 
\section{Acknowledgement}
L.~T. would like to thank the organizers of the  HADRON 2021 Conference for the opportunity to give this parallel talk.
L.~T. and Y.~H. are supported by National Research Foundation (NRF) funded by the Ministry of Science of Korea with Grant No.~2021R1F1A1061387(LT) and  2020R1F1A1076267(YH).
N. H. is supported by Department of Atomic Energy (DAE), India via National Institute of Science Education and Research.
\end{multicols}

\medline
\begin{multicols}{2}
%

\providecommand{\href}[2]{#2}\begingroup\raggedright\endgroup

\end{multicols}
\end{document}